\documentclass[11pt]{iopart}

\usepackage{iopams}
\usepackage{setstack}
\usepackage{mathrsfs}
\usepackage[square,sort&compress,numbers]{natbib}
\usepackage[colorlinks=true,citecolor=blue,urlcolor=green]{hyperref}
\usepackage{perpage}
\MakePerPage{footnote}

\hypersetup{
    pdftitle={Li(e)nearity}, 
    pdfauthor={R. Leone and F. Haas},   
    pdfcreator={R. Leone and F. Haas},   
    pdfkeywords={linearity}{Lie and Noether symmetries}{Jacobi last multiplier}{higher-order Lagrangian system}	
}

\makeatletter
\renewcommand{\dddot}[1]{%
  {\mathop{\kern\z@#1}\limits^{\vbox to-1.4\ex@{\kern-\tw@\ex@
   \hbox{\normalfont ...}\vss}}}}
\renewcommand{\ddddot}[1]{%
  {\mathop{\kern\z@#1}\limits^{\vbox to-1.4\ex@{\kern-\tw@\ex@
   \hbox{\normalfont....}\vss}}}}
\makeatother

\begin{document}

\title[Li(e)nearity]{Li(e)nearity}

\author{Rapha\"{e}l Leone$^{1,2}$ and Fernando Haas$^3$}

\address{$^1$ Groupe  de  Physique  Statistique, IJL, UMR CNRS 7198, Universit\'e  de  Lorraine, 54506 Vand\oe uvre-l\`es-Nancy,  France}
\address{$^2$ Laboratoire d'Histoire des Sciences et de Philosophie -- Archives Henri Poincar\'e, UMR CNRS 7117, Universit\'e de Lorraine, 54501 Nancy, France}
\address{$^3$ Instituto de F\'{\i}sica, Universidade Federal do Rio Grande do Sul, Avenida Bento Gon\c{c}alves 9500, 91501-970 Porto Alegre, RS, Brasil}

\ead{\href{mailto:raphael.leone@univ-lorraine.fr}{raphael.leone@univ-lorraine.fr}, \href{fernando.haas@ufrgs.br}{fernando.haas@ufrgs.br}} 

\begin{abstract}
We demonstrate the fact that linearity is a meaningful symmetry in the sense of Lie and Noether. The role played by that `linearity symmetry' in the quadrature of linear ordinary second-order differential equations is reviewed, by the use of canonical coordinates and the identification of a Wronskian-like conserved quantity as Lie invariant. The Jacobi last multiplier associated with two independent linearity symmetries is applied to derive the Caldirola-Kanai Lagrangian from symmetry principles. Then the symmetry is recognized to be also a Noether one. Finally, the study is extended to higher-order linear ordinary differential equations, derivable or not from an action principle.
\end{abstract}

\pacs{02.30.Hq,45.20.Jj,45.50.Dd}
\vspace{2pc}
\noindent{\it Keywords\/}: linearity, Lie and Noether symmetries, Jacobi last multiplier, higher-order Lagrangian system

\maketitle

\section{Introduction}

Let us consider the most general second-order linear differential equation (LDE), with independent variable $t$ and dependent one $q$, put in the standard form
\begin{equation}
\Delta_2(t,q,\dot q,\ddot q)=\ddot q+a(t)\dot q+b(t)q+c(t)=0,\label{ODE}
\end{equation} 
where the overdot denotes differentiation with respect to $t$. In classical mechanics, such an equation describes a driven damped harmonic oscillator with \textit{a priori} time-dependent frequency, dissipation rate and excitation. In undergraduate textbooks on mathematics~\cite{r1,r2,r3}, one learns that once a nonzero solution $s(t)$ of the homogeneous equation
\begin{equation}
\Delta_{2\rm h}(t,q,\dot q,\ddot q)=\ddot q+a(t)\dot q+b(t)q=0\label{homogeneous}
\end{equation}
is known, the differential equation may be reduced to a first-order one in the derivative of the dependent variable $z=q/s(t)$. The linearity\footnote{The adjective `linear' frequently used to designate a differential equation such as~\eref{ODE} is somewhat regrettable in the inhomogeneous case where $c(t)\ne0$. Indeed, although the independent variable $q$ and its derivatives appear linearly in~\eref{ODE}, the solution space is not linear but affine and it would have been preferable to speak in terms of affine differential equations. We have chosen to follow the accepted terminology.} of equation~\eref{ODE} is the key-ingredient of the validity of this traditional recipe. However, the introduction of the variable $z$ is commonly presented as a lucky and hence unsatisfactory ansatz. It overlooks the symmetry origin of the method of reduction, as was observed by Sophus Lie himself, the father of the theory of continuous transformation groups. In his classical book entitled \textit{Vorlesungen \"uber Differentialgleichungen mit bekannten infinitesimalen Transformationen}~\cite{Lie1891}, he details some examples of applications of his theory to differential equations for their reduction. In particular, he treats the case of equation~\eref{ODE} and exhibits the symmetry responsible of its reduction, a symmetry stemming only from the linearity property. Actually, in place of $z$, he uses the more convenient variable $w=\dot qs(t)-q\dot s(t)$ that will be named `Wronskian variable' in this article\footnote{Our notations differ from Lie's ones in reference~\cite{Lie1891}. To obtain the latter, make the replacements $t\to x$, $q\to y$, $s(t)\to z(x)$, $w\to v$.}.

Evidently, since the seminal works of Lie, the aforesaid `linearity symmetry' has been well recognized in the long history concerning the general second-order LDE, where it has been treated mainly as a marginal result. An exception is reference~\cite{Kumei}, where the linearity symmetry is used to the order reduction and quadrature of homogeneous second-order LDEs. The existence of conserved quantities --- linear in velocities --- for linear time-dependent systems and its connection to invariance principles has been obtained in the literature~\cite{Prince, Leach1, Ray1, Ray2, Pedrosa, Soliani, Aguirre}. Furthermore, the Noetherian character of the Wronskian has been recognized~\cite{Castanos1, Qin}, at least in the case of the time-dependent harmonic oscillator where $a(t) = c(t) = 0$ in~\eref{ODE}. Generalisations to linear in velocity invariants for arbitrary multidimensional quadratic Hamiltonian and Lagrangian systems have been also provided~\cite{Castanos2}. More frequently, there is the emphasis~\cite{Lutzky} put on the quadratic invariants of the Ermakov-Lewis class~\cite{Ermakov, Lewis, RayReid, Mancas}, also known as Courant-Snyder invariants in the context of accelerator physics~\cite{Qin, Courant}.  

If the above is possibly not an exhaustive list, it appears to us that the central role played by the linearity symmetry is underestimated in the literature. The \textit{raison d'\^etre} of the present article is to show that this symmetry is at the source of all the well-known mathematical techniques associated with LDEs in terms of integrating factors, Wronskian considerations, Abel's identity, Caldirola-Kanai Lagrangian, and so forth. Actually, as will be emphasized, it is more than simply a Lie symmetry since it leaves invariant the defining function $\Delta_2$ and not only the differential equation $\Delta_2=0$. That property has the great advantage of allowing a simplified treatment of the symmetry without the need for the usual machinery of Lie generators. Our purpose is also to generalise the results obtained for second-order LDEs to higher-order ones.  

The paper is organized as follows. In section~\ref{sec:Lie}, after some considerations on first-order LDEs, the Lie symmetry associated with the linearity of~\eref{ODE} and the induced first integral are fully discussed. Equation~\eref{ODE} being known to be derivable from the celebrated Lagrangian named after Caldirola~\cite{Caldirola} and Kanai~\cite{Kanai}, section~\ref{sec:Noether} starts with its reconstruction based on the linearity symmetry, through the Jacobi last multiplier~\cite{Jacobi1844,Jacobi1845} generated by the symmetry~\cite{Lie1874b,Whittaker}. Then, the symmetry is shown to be a Noether one of the Lagrangian, in accordance with previous findings~\cite{Castanos1, Qin}. In Section~\ref{sec:higher}, we consider the extension of the linearity symmetry to third-order LDEs as well as fourth-order ones derivable from an action principle. Finally, section~\ref{sec:generalisation} is devoted to a complete generalisation at arbitrary orders.  
 
\section{Linearity as a Lie point symmetry}\label{sec:Lie}

\subsection{The preliminary case of first-order linear differential equations}\label{preliminary}

Before considering the second-order LDE~\eref{ODE}, let us first have a look at the first-order one whose general form is
\begin{equation}
\Delta_1(t,q,\dot q)= \dot q+a(t)q+b(t)=0.\label{ODE1}
\end{equation}
If $s(t)$ is a nonzero solution of the homogeneous equation
\begin{equation}
\Delta_{1\rm h}(t,q,\dot q)= \dot q+a(t)q=0,\label{ODE1h}
\end{equation}
then the transformation $(t,q)\to(t,q+s(t))$ leaves $\Delta_1$ invariant in the sense that
\begin{equation*}
\Delta_1(t,q+s(t),\dot q+\dot s(t))=\Delta_1(t,q,\dot q)+\Delta_{1\rm h}(t,s(t),\dot s(t))=\Delta_1(t,q,\dot q).
\end{equation*}
Alternatively stated, it is a finite symmetry of $\Delta_1$. By linearity, the function $\varepsilon s(t)$ remains a solution of~\eref{ODE1h} for any value of a real parameter $\varepsilon$ thus the family of transformations
\begin{equation}
\mathscr L(\varepsilon)\colon\qquad (t,q)\longrightarrow(t,q+\varepsilon s(t))\label{transformation}
\end{equation}
is a continuous one-parameter symmetry group of $\Delta_1$. It leaves \textit{a fortiori} $\Delta_1$ invariant on-shell, i.e. it is a Lie point symmetry of equation~\eref{ODE1}. Moreover, the time $t$ is left unaffected while $z=q/s(t)$ merely undergoes a translation by $\varepsilon$:
\begin{equation*}
z=\frac{q}{s(t)}\longrightarrow\frac{q+\varepsilon s(t)}{s(t)}=z+\varepsilon.
\end{equation*}
Thus, $(t,z)$ is a couple of variables for which the continuous transformation reduces to
\begin{equation*}
\mathscr L(\varepsilon)\colon\qquad (t,z)\longrightarrow(t,z+\varepsilon).
\end{equation*}
These so-called canonical variables of $\mathscr L(\varepsilon)$ have the great advantage of leaving the derivatives of $z$ unchanged. Let us exploit this property by first expressing $\Delta_1$ in terms of the new variables through
\begin{equation*}
\Delta'_1(t,z,\dot z)=\Delta_1(t,q,\dot q)=\Delta_1(t,s(t)z,s(t)\dot z+\dot s(t)z).
\end{equation*}
Equation~\eref{ODE1} is equivalent to $\Delta'_1=0$ and the invariance under $\mathscr L(\varepsilon)$ reads now
\begin{equation*}
\Delta'_1(t,z+\varepsilon,\dot z)=\Delta'_1(t,z,\dot z).
\end{equation*}
This equality simply means that $\Delta'_1$ does not depend on $z$ and that the original LDE~\eref{ODE1} is reduced to an equation in $t$ and $\dot z$ only. Explicitly:
\begin{equation*}
\Delta'_1=s(t)\dot z+b(t)=0.
\end{equation*}
Being linear in $\dot z$, this equation is easily integrated to provide the general solution $q(t)$:
\begin{equation}
q(t)=s(t)\bigg[C-\int\frac{b(t)}{s(t)}\,\rmd t\bigg]=\rme^{-\int a(t)\rmd t}\bigg[C-\int b(t)\,e^{\int a(t)\rmd t}\rmd t\bigg],\label{solution1}
\end{equation}
where $C$ is a constant of integration and where one has specified $s(t)$ to be the obvious solution $\exp(-\int a(t)\rmd t)$ of~\eref{ODE1h}. Hence, thanks to the linearity symmetry, equation~\eref{ODE1} is solved by a single quadrature. This fact was seen and discussed by Lie as an application of his theory to first-order differential equations, at the end of chapter 8 in reference~\cite{Lie1891}. 

Actually, the most important result regarding Lie's theory in the realm of first-order differential equations is certainly the association of an integrating factor (also named Euler's multiplier) with any Lie point symmetry~\cite{Lie1874a}. In reference~\cite{Lie1891}, this association is stated and made explicit in chapter 6, theorem 8. With our notations, it can be restated as follows: if a first-order differential equation, put in form
\begin{equation}
T(t,q)\dot q-Q(t,q)=0,\label{1order}
\end{equation}
is invariant under a transformation 
\begin{equation*}
(t,q)\longrightarrow (t+\varepsilon\,\eta(t,q),q+\varepsilon\,\xi(t,q))
\end{equation*}
then it admits as integrating factor
\begin{equation*}
\mu(t,q)=\frac{1}{T\,\eta-Q\xi}\,.
\end{equation*}
In other words, there exists some function $I(t,q)$ such that
\begin{equation*}
\mu(t,q)\big[T(t,q)\dot q-Q(t,q)\big]=\dot I.
\end{equation*}
Clearly, $I$ keeps a constant value $C$ along the solutions and the equality $I=C$ is said to be a first integral of equation~\eref{1order}. In the case of the LDE~\eref{ODE1}, the linearity symmetry~\eref{transformation} gives rise to the integrating factor $1/s(t)=\exp(\int a(t)\rmd t)$ and to the first integral
\begin{equation}
I(t,q)= \frac{q}{s(t)}+\int\frac{b(t)}{s(t)}\,\rmd t=q\,\rme^{\int a(t)\rmd t}+\int b(t)\,\rme^{\int a(t)\rmd t}\rmd t=C\label{FI1}
\end{equation}
which amounts to~\eref{solution1}. Note that $I$ transforms like $z$ under $\mathscr L(\varepsilon)$. If one had luckily chosen $I$ instead of $z$ as canonical variable, the reduced equation would have been simply $\dot I=0$.

In conclusion, the linearity symmetry~\eref{transformation} is the source of the well-known integrating factor $\exp(\int a(t)\rmd t)$ of equation~\eref{ODE1}. It is worth noting \textit{en passant} that an homogeneous LDE such as~\eref{homogeneous} or~\eref{ODE1h}, whose solution space is linear, is evidently also invariant under a rescaling of $q$ (here, the scale invariance is a Lie symmetry of the LDE and not merely an invariance of its left-hand side). However, we will not be concerned with this eventual extra symmetry and will remain focused on the invariance under the addition of solutions of the associated homogeneous LDE.

\subsection{The case of second-order differential equations}

Now let us consider the second-order LDE~\eref{ODE} and let $s(t)$ be a nonzero solution of the homogeneous equation~\eref{homogeneous}. Here again, the continuous transformation
\begin{equation*}
\mathscr L(\varepsilon)\colon\qquad (t,q)\longrightarrow(t,q+\varepsilon s(t))
\end{equation*}
leaves $\Delta_2$ invariant. Exactly for the same reasons as in the previous paragraph, the expression of~\eref{ODE}, in terms of the canonical variables $t$ and $z=q/s(t)$, becomes a first-order LDE in $\dot z$. Explicitly:
\begin{equation*}
\Delta'_2=s(t)\ddot z+\big[a(t)s(t)+2\dot s(t)\big]\dot z+c(t).
\end{equation*}
However, the reduced LDE takes a simpler form if one introduces the Wronskian variable
\begin{equation*}
w=\dot q s(t)-q\dot s(t)=s^2(t)\dot z
\end{equation*}
which is also a first-order invariant of $\mathscr L(\varepsilon)$. Indeed, it becomes
\begin{equation*}
\dot w+a(t)w+c(t)s(t)=0.
\end{equation*}
Applying formula~\eref{FI1} to that LDE, one obtains directly the first integral
\begin{equation}
I(t,q,\dot q)=\Big(\dot q s(t)-q\dot s(t)\Big)\rme^{\int a(t)\rmd t}+\int c(t)s(t)\,\rme^{\int a(t)\rmd t}\rmd t=C.\label{FI}
\end{equation}
It is by itself an invariant of $\mathscr L(\varepsilon)$. In the homogeneous case where $c(t)=0$, equation~\eref{FI} is nothing else but Abel's identity in which the exponential factor exactly compensates the amplitude damping of both $q(t)$ and $s(t)$. Substituting $s^2(t)\dot z$ for $w$ in~\eref{FI} yields an expression of $\dot z$ as a function of $t$ which can easily be integrated to give the general solution $q(t)$ of~\eref{ODE}, \textit{videlicet}
\begin{equation*}
q(t)=s(t)\left\{\int\frac{1}{s^2(t)}\left[C-\int c(t)s(t)\,\rme^{\int a(t)\rmd t}\rmd t\right]\rme^{-\int a(t)\rmd t}\rmd t+C'\right\},
\end{equation*}
where $C'$ is another constant of integration.  

Since the solution space of the homogeneous LDE~\eref{homogeneous} is a two-dimensional vector space, the whole group associated with the linearity symmetry is actually the two-parameter symmetry group 
\begin{equation*}
\mathscr L(\varepsilon_1,\varepsilon_2)\colon\qquad(t,q)\longrightarrow(t,q+\varepsilon_1s_1(t)+\varepsilon_2s_2(t)),
\end{equation*}
where $s_1(t)$ and $s_2(t)$ are two independent solutions of~\eref{homogeneous}. The knowledge of $s_1(t)$ and $s_2(t)$ enables an algebraic resolution of~\eref{ODE} which constitutes an alternative to the usual method of variation of the parameters. Indeed, they induce respectively two first integrals $I_1(t,q,\dot q)=C_1$ and $I_2(t,q,\dot q)=C_2$ as given by formula~\eref{FI}. They form a Cramer's system in $q$ and $\dot q$ from which $q(t)$ can be extracted. 

We end this subsection with the couples of first integrals generated by the linearity symmetry in the two most relevant examples encountered in physics, at the undergraduate level.

\subsubsection{Example 1: the harmonic oscillator}

Let us consider the equation of motion of the harmonic oscillator
\begin{equation}
\ddot q+\omega_0^2q=0\label{OH},
\end{equation}
where the natural frequency $\omega_0$ is a constant. Here, the equation is homogeneous and the formal first integral~\eref{FI} is the Wronskian itself:
\begin{equation*}
I(t,q,\dot q)=\dot q s(t)-q\dot s(t).
\end{equation*}
Since two independent solutions of~\eref{OH} are $s_1(t)=\cos(\omega_0 t)$ and $s_2(t)=\sin(\omega_0t)$, the linearity symmetry generates the two independent first integrals
\begin{equation*}
I_1=\dot q\cos(\omega_0t)+\omega_0 q\sin(\omega_0t)\qquad{\rm and}\qquad I_2=\dot q\sin(\omega_0t)-\omega_0 q\cos(\omega_0t).
\end{equation*}

\subsubsection{Example 2: the driven damped harmonic oscillator}

Let us now move on to the equation of motion of a damped harmonic oscillator driven by a sinusoidal excitation force:
\begin{equation*}
\ddot q+2\gamma\dot q+\omega_0^2q-F\cos(\omega_{\rm e} t)=0,
\end{equation*} 
where the natural and excitation frequencies, $\omega_0$ and $\omega_{\rm e}$, are constants as well as the dissipation rate $\gamma$ and the characteristic force $F$. The formal first integral is
\begin{equation*}
I(t,q,\dot q)=\Big(\dot q s(t)-q\dot s(t)\Big)\rme^{2\gamma t}-F\int\cos(\omega_{\rm e} t)s(t)\,\rme^{2\gamma t}\rmd t.
\end{equation*}
Suppose that we are in the underdamped regime in which case $s_1(t)=\cos(\omega t)\rme^{-\gamma t}$ and $s_2(t)=\sin(\omega t)\rme^{-\gamma t}$ are two real independent solutions of the homogeneous equation, where $\omega=(\omega_0^2-\gamma^2)^{1/2}$. The corresponding first integrals are
\begin{eqnarray*}
\fl I_1=\bigg[\dot q\cos(\omega t)+q\big(\gamma\cos(\omega t)+\omega\sin(\omega t)\big) 
-\frac{F \sin((\omega + \omega_{\rm e}) t+\beta_+)}{2\sqrt{(\omega+\omega_{\rm e})^2+\gamma^2}} -\frac{F \sin((\omega - \omega_{\rm e}) t+\beta_-)}{2\sqrt{(\omega-\omega_{\rm e})^2+\gamma^2}}\bigg]\rme^{\gamma t},\\
\fl I_2=\bigg[\dot q\sin(\omega t)+q\big(\gamma\sin(\omega t)-\omega\cos(\omega t)\big)+\frac{F \cos((\omega + \omega_{\rm e}) t+\beta_+)}{2\sqrt{(\omega+\omega_{\rm e})^2+\gamma^2}} +\frac{F \cos((\omega - \omega_{\rm e}) t+\beta_-)}{2\sqrt{(\omega-\omega_{\rm e})^2+\gamma^2}}\bigg]\rme^{\gamma t},
\end{eqnarray*}
where one has introduced the angles
\begin{equation*}
\beta_\pm=\arctan\bigg(\frac{\gamma}{\omega\pm\omega_{\rm e}}\bigg).
\end{equation*}

\section{Linearity as a Noether point symmetry}\label{sec:Noether}

The equation of motion~\eref{ODE} is known to be equivalent to the Euler-Lagrange equation derived from the Caldirola-Kanai Lagrangian~\cite{Caldirola, Kanai}
\begin{equation}
L(t,q,\dot q)=\bigg(\frac12\,\dot q^2-\frac12\,b(t)q^2-c(t)q\bigg)\rme^{\int a(t)\rmd t}\,.\label{Lagrangian}
\end{equation}
A first manner of constructing this Lagrangian is to perform the point transformation $q\to Q=\exp(\int a(t)\rmd t/2)q$ in equation~\eref{ODE}. It maps the initial problem onto a problem of a non dissipative oscillator with amplitude $Q$ governed by the equation
\begin{equation*}
\ddot Q+\bigg[b(t)-\frac12\, \dot a(t)-\frac{1}{4}\,a^2(t)\bigg]Q+c(t)\,\rme^{\frac12\int a(t)\rmd t}=0
\end{equation*}
derivable from the standard Lagrangian
\begin{equation}
L=\frac12\,\dot Q^2-\frac12\bigg[b(t)-\frac12\, \dot a(t) -\frac{1}{4}\,a^2(t)\bigg]Q^2-c(t)\,\rme^{\frac12\int a(t)\rmd t}Q.\label{LagrangianQ}
\end{equation}
Then, performing the inverse transformation $Q\to q$ in~\eref{LagrangianQ} provides
\begin{equation*}
L=\bigg(\frac12\,\dot q^2-\frac12\,b(t)q^2-c(t)q\bigg)\rme^{\int a(t)\rmd t}+\frac{\rmd}{\rmd t}\bigg(\frac14\,a(t)q^2\rme^{\int a(t)\rmd t}\bigg)
\end{equation*}
and one has thereby obtained~\eref{Lagrangian} up to a removable total derivative. However, this result may be derived without any guess, only by the application of the linearity symmetry in conjunction with the theory of Jacobi multipliers~\cite{Jacobi1844,Jacobi1845,Lie1891,Whittaker} which generalise Euler's ones. It is well-known since the works of Lie~\cite{Lie1874b} that if a second-order differential equation
\begin{equation}
\ddot q-F(t,q,\dot q)=0\label{Newton}
\end{equation}
possesses two independent Lie point symmetries
\begin{equation*}
(t,q)\longrightarrow(t+\varepsilon\,\xi_i(t,q),q+\varepsilon\,\eta_i(t,q))\qquad(i=1,2)
\end{equation*}
then the quantity
\begin{equation*}
M(t,q,\dot q)=\left|\begin{array}{ccc}
1 & \dot q & F\\
\xi_1 & \eta_1 & \dot\eta_1-\dot q\dot\xi_1\\
\xi_2 & \eta_2 & \dot\eta_2-\dot q\dot\xi_2\\
\end{array}\right|^{-1}
\end{equation*}
is such that
\begin{equation*}
\frac{\partial}{\partial\dot q}(MF)+\frac{\partial}{\partial q}(M\dot q)+\frac{\partial M}{\partial t}=0.
\end{equation*}
It is a so-called Jacobi last multiplier of equation~\eref{Newton}, see also Refs.~\cite{Lie1891} (chap.~15, \S~5, theorem~32) and~\cite{Whittaker}. What is important for our purpose is that, beyond their profound signification, Jacobi last multipliers bring solutions to the inverse Lagrange problem in the case of a single coordinate. Indeed, a last multiplier $M$ guarantees the existence of a Lagrangian $L$, constrained by~\cite{Whittaker}
\begin{equation*}
\frac{\partial^2 L}{\partial\dot q^2}=M,
\end{equation*}
whose Euler-Lagrange equation amounts to equation~\eref{Newton}. Using the linearity symmetries generated by $s_1(t)$ and $s_2(t)$, equation~\eref{ODE} admits the Jacobi last multiplier
\begin{equation*}
M=\Bigg|\begin{array}{cc}
s_1(t) & \dot s_1(t)\\
s_2(t) & \dot s_2(t)
\end{array}\Bigg|^{-1}=\frac{1}{ s_1(t)\dot s_2(t)-\dot s_1(t) s_2(t)}=K\,\rme^{\int a(t)\rmd t},
\end{equation*}
thanks to Abel's identity. Here, $K$ is a nonzero constant without any signification which will be set to unity after multiplying $s_1(t)$ or $s_2(t)$ by a constant factor if necessary. Interestingly enough, the expressions of $s_1(t)$ and $s_2(t)$ are not required to infer the last multiplier, the knowledge of their existence suffices. Then, the Lagrangian has \textit{a priori} the form
\begin{equation*}
L(t,q,\dot q)=\Bigg(\frac12\,\dot q^2+f_1(t,q)\dot q+f_0(t,q)\Bigg)\rme^{\int a(t)\rmd t}.
\end{equation*}
However one can always remove a term linear in $\dot q$ by a gauge transformation, that is, by adding the total derivative of a suitable function of $(t,q)$. Thus, without lost in generality, one can make the gauge choice $f_1(t,q)=0$ yielding the Euler-Lagrange equation
\begin{equation*}
\mathsf E(L)=\frac{\partial L}{\partial q}-\frac{\rmd}{\rmd t}\frac{\partial L}{\partial\dot q}=-\Bigg(\ddot q+a(t)\dot q-\frac{\partial f_0}{\partial q}(t,q)\Bigg)\rme^{\int a(t)\rmd t}.
\end{equation*}
It is equivalent to~\eref{ODE} if
\begin{equation*}
f_0(t,q)=-\frac12\,b(t)q^2-c(t)q
\end{equation*}
and one has re-obtained the Lagrangian\footnote{We point out that, whatever the potential $V(t,q)$ be, a dynamic equation $\ddot q+a(t)\dot q+\partial_qV(t,q)$ is always deducible from the Caldirola-Kanai Lagrangian
\begin{equation*}
L=\Big(\frac12\,\dot q^2-V(t,q)\Big)\rme^{\int a(t)\rmd t}.
\end{equation*}
This is because $M=\rme^{\int a(t)\rmd t}$ is actually an universal last multiplier of the equation of motion. However, it is \textit{a priori} not a consequence of any symmetry (one can find in reference~\cite{Gourieux} the list of potentials for which the above dynamic equation admits a point symmetry). Interestingly, for linear equations, the existence of the last multiplier ceases to be `accidental'.}~\eref{Lagrangian}. Under the transformation~\eref{transformation}, it becomes
\begin{equation*}
L(t,q+\varepsilon s(t),\dot q+\varepsilon\dot s(t))=L(t,q,\dot q)+\varepsilon\Big(\dot q\dot s(t)-\big[b(t)q+c(t)\big] s(t)\Big) \rme^{\int a(t)\rmd t}
\end{equation*}
up to the first order in $\varepsilon$. Hence, using the fact that $s(t)$ is a solution of~\eref{homogeneous}, one has
\begin{equation}
\delta L=L(t,q+\varepsilon s(t),\dot q+\varepsilon\dot s(t))-L(t,q,\dot q)=\varepsilon\,\frac{\rmd f}{\rmd t}\,,\label{one}
\end{equation}
where
\begin{equation*}
f=q\dot s(t)\,\rme^{\int a(t)\rmd t}-\int c(t)s(t)\,\rme^{\int a(t)\rmd t}\rmd t.
\end{equation*}
Since $L$ merely undergoes a gauge transformation under~\eref{transformation}, the later is a Noether point symmetry. On the other hand, one has independently on the form of $L$:
\begin{eqnarray}
\delta L&=\varepsilon\Bigg[s(t)\frac{\partial L}{\partial q}+\dot s(t)\frac{\partial L}{\partial\dot q}\Bigg]=\varepsilon\Bigg[s(t)\mathsf E(L)+\frac{\rmd}{\rmd t}\Bigg(s(t)\frac{\partial L}{\partial\dot q}\Bigg)\Bigg],\label{two}
\end{eqnarray}
up to the first order in $\varepsilon$. Equations~\eref{one} and~\eref{two} give 
\begin{equation*}
\frac{\rmd}{\rmd t}\Bigg[s(t)\frac{\partial L}{\partial\dot q}-f(t,q)\Bigg]=-s(t)\mathsf E(L).
\end{equation*}
Along the solutions $q(t)$, the right-hand side vanishes so the symmetry generates the conservation of the expression in brackets, namely the Noether invariant which coincide with Lie's one~\eref{FI}.

\section{Higher-order linear differential equations}\label{sec:higher}

\subsection{Linearity symmetries of third-order linear differential equations}

One can wonder about the extension of the linearity symmetry and the associated Wronskian-type conservation law to higher-order LDEs. For this purpose and for the sake of illustration, we now consider a third-order LDE 
\begin{equation}
\Delta_3(t,q,\dot q,\ddot q,\dddot q)=\dddot q+a(t)\ddot q+b(t)\dot q+c(t)q+d(t)=0.\label{ODE3}
\end{equation}
Let $s_1(t)$ be a nonzero solution of the homogeneous equation
\begin{equation}
\Delta_{3\rm h}(t,q,\dot q,\ddot q,\dddot q)=\dddot q+a(t)\ddot q+b(t)\dot q+c(t)q=0.\label{ODE3h}
\end{equation}
Yet again, the transformation 
\begin{equation*}
\mathscr L_1(\varepsilon)\colon\qquad (t,q)\longrightarrow(t,q+\varepsilon s_1(t))
\end{equation*}
is a symmetry of $\Delta_3$. Introducing the canonical variable $z_1=q/s_1(t)$, equation~\eref{ODE3} is reduced to a second-order LDE
\begin{equation*}
\Delta'_3(t,\dot z_1,\ddot z_1,\dddot z_1)=\Delta_3(t,q,\dot q,\ddot q,\dddot q)=0
\end{equation*}
whose dependent variable is the invariant
\begin{equation*}
\dot z_1=\frac{\rmd}{\rmd t}\bigg(\frac{q}{s_1(t)}\bigg)=\frac{1}{s_1^2(t)}\Bigg|\begin{array}{cc}
s_1(t) & q\\ \dot s_1(t) & \dot q
\end{array}\Bigg|
\end{equation*}
of $\mathscr L_1(\varepsilon)$. Now, let $s_2(t)$ be another independent solution of~\eref{ODE3h}. By construction, $\Delta'_3$ inherits the invariance under the symmetry group
\begin{equation*}
\mathscr L_2(\varepsilon)\colon\qquad (t,q)\longrightarrow(t,q+\varepsilon s_2(t)).
\end{equation*}
The action of $\mathscr L_2(\varepsilon)$ on $\dot z_1$ is simply
\begin{equation*}
\dot z_1\longrightarrow\frac{1}{s_1^2(t)}\Bigg|\begin{array}{cc}
s_1(t) & q+\varepsilon s_2(t)\\ \dot s_1(t) & \dot q+\varepsilon\dot s_2(t)
\end{array}\Bigg|=\dot z_1+\frac{\varepsilon}{s_1^2(t)}\Bigg|\begin{array}{cc}
s_1(t) & s_2(t)\\ \dot s_1(t) & \dot s_2(t)
\end{array}\Bigg|.
\end{equation*}
Hence, $\mathscr L_2(\varepsilon)$ merely translates by $\varepsilon$ the canonical variable
\begin{equation*}
z_2=\frac{s^2_1(t)\dot z_1}{\Bigg|\begin{array}{cc}
s_1(t) & s_2(t)\\ \dot s_1(t) & \dot s_2(t)
\end{array}\Bigg|}=\frac{\Bigg|\begin{array}{cc}
s_1(t) & q\\ \dot s_1(t) & \dot q
\end{array}\Bigg|}{\Bigg|\begin{array}{cc}
s_1(t) & s_2(t)\\ \dot s_1(t) & \dot s_2(t)
\end{array}\Bigg|}
\end{equation*}
and one obtains a first-order LDE in
\begin{equation*}
\dot z_2=\frac{s_1(t)w}{\Bigg|\begin{array}{cc}
s_1(t) & s_2(t)\\ \dot s_1(t) & \dot s_2(t)
\end{array}\Bigg|^2}\qquad{\rm where}\qquad w=\left|\begin{array}{ccc}
s_1(t) & s_2(t) & q \\ \dot s_1(t) & \dot s_2(t) & \dot q \\ \ddot s_1(t) & \ddot s_2(t) & \ddot q
\end{array}\right|,
\end{equation*}
or in the Wronskian variable $w$ itself. It may be deduced directly by deriving $w$. Since each row is the derivative of the preceding, one has
\begin{equation*}
\dot w=\left|\begin{array}{ccc}
s_1(t) & s_2(t) & q \\ \dot s_1(t) & \dot s_2(t) & \dot q \\ \dddot s_1(t) & \dddot s_2(t) & \dddot q
\end{array}\right|=-a(t)w-d(t)\Bigg|\begin{array}{cc}
s_1(t) & s_2(t)\\ \dot s_1(t) & \dot s_2(t)
\end{array}\Bigg|.
\end{equation*}
It is easily integrated to yield the first integral
\begin{equation}
\fl I_3(t,q,\dot q,\ddot q)=\left|\begin{array}{ccc}
s_1(t) & s_2(t) & q \\ \dot s_1(t) & \dot s_2(t) & \dot q \\ \ddot s_1(t) & \ddot s_2(t) & \ddot q\end{array}
\right|\,\rme^{\int a(t)\rmd t}+\int d(t)\Bigg|\begin{array}{cc}
s_1(t) & s_2(t)\\ \dot s_1(t) & \dot s_2(t)\end{array}\Bigg|\,\rme^{\int a(t)\rmd t}\rmd t=C_3\label{I3}
\end{equation}
which, like $\dot z_2$ or $w$, is a simultaneous Lie invariant of $\mathscr L_1(\varepsilon)$ and $\mathscr L_2(\varepsilon)$. The last equality may be integrated in $z_2$, then in $z_1$ to provide the general expression of $q(t)$. However, things becomes easier when one considers the whole linearity symmetry group through the introduction of a third independent solution $s_3(t)$ of~\eref{ODE3h}. Then, permuting cyclically the indices $1,2,3$ in~\eref{I3} yields two other first integrals $I_1=C_1$ and $I_2=C_2$. The three first integrals define a Cramer system in $q,\dot q,\ddot q$ from which $q(t)$ can be extracted.

\subsection{Linearity symmetries of third-order Lagrangians}

We now consider a Lagrangian depending also on the acceleration, so that $L = L(t,q,\dot{q},\ddot{q})$. In this case~\cite{Miron, Leon} the Euler-Lagrange equation reads
\begin{equation*}
\mathsf E(L)=\frac{\partial L}{\partial q}-\frac{\rmd}{\rmd t}\frac{\partial L}{\partial\dot q}+\frac{\rmd^2}{\rmd t^2}\frac{\partial L}{\partial\ddot q} = 0 \,.
\end{equation*}
It is certainly linear if the Lagrangian has the form
\begin{equation}
L(t,q,\dot q,\ddot q)=\frac12\,A(t)\,\ddot q^2+\frac12\,B(t)\,\dot q^2+\frac12\,C(t)\,q^2-D(t)\,q.\label{lag}
\end{equation}
Notice that further monomials in $q\ddot q, \dot q\ddot q, q\dot q, \ddot q, \dot q$ would be superfluous. Indeed, they can be accommodated in the above picture by repeating Leibniz' product rule as many time as necessary and taking into account that total derivatives do not contribute to the Euler-Lagrange equations. For instance, one has 
\begin{eqnarray*}
E(t)q\ddot q&=\frac{\rmd}{\rmd t}\Big(E(t) q\dot q\Big)-\dot E(t)q\dot q-E(t)\dot q^2\\
&=\frac{\rmd}{\rmd t}\bigg(E(t) q \dot q - \frac12\,\dot E(t) q^2\bigg)+\frac12\,\ddot E(t) q^2-E(t)\dot q^2,
\end{eqnarray*}
and the relevant terms in the second line are seen to fit with~\eref{lag}. The Euler-Lagrange equation for~\eref{lag} reads
\begin{equation*}
A(t) \ddddot q
+ 2 \dot A(t)\dddot q + \big[\ddot A(t) - B(t)\big]\ddot q - \dot B(t)\dot q + C(t) q - D(t) =0.
\end{equation*}
We point out that not all fourth-order LDE fulfils the above equation, so that the variational principle imposes some restrictions. We are not aware of a suitable Lagrangian for the general fourth-order LDE. In addition, observe that for $A(t)= 0$ one goes directly from a fourth-order to a second-order equation in this case. 

Now, let $s(t)$ be a nonzero solution of the associated homogeneous equation:
\begin{equation*}
A(t) \ddddot q
+ 2 \dot A(t)\dddot q + \big[\ddot A(t) - B(t)\big]\ddot q - \dot B(t)\dot q + C(t) q =0.
\end{equation*}
Using this property, one verifies that the transformation $q\to q+\varepsilon s(t)$ is a Noether point symmetry of $L$ since
\begin{equation}
\delta L=L(t,q+\varepsilon s(t),\dot q+\varepsilon\dot s(t),\ddot q+\varepsilon\ddot s(t))-L(t,q,\dot q,\ddot q)=\varepsilon\,\frac{\rmd f}{\rmd t}\,,\label{deltaL3s}
\end{equation} 
with
\begin{equation*}
f = A(t)\ddot s(t)\dot q - A(t)\dddot s(t)q - \dot A(t)\ddot s(t) q+ B(t)\dot s(t) q - \int D(t)s(t)\rmd t.
\end{equation*}
However, independently of the form of $L$, the action of the transformation is generally
\begin{eqnarray}
\delta L&=\varepsilon\bigg[s(t)\frac{\partial L}{\partial q} + \dot s(t)\frac{\partial L}{\partial\dot q} + \ddot s(t)\frac{\partial L}{\partial\ddot q}\bigg]\nonumber\\ 
&=\varepsilon\bigg[s(t)\mathsf E(L) + \frac{\rmd}{\rmd t}\bigg(\dot s(t) \frac{\partial L}{\partial\ddot q} - s(t) \frac{\rmd}{\rmd t}\frac{\partial L}{\partial\ddot q}+ s(t)\frac{\partial L}{\partial\dot q}\bigg)\bigg].\label{deltaL3}
\end{eqnarray}
Then, one concludes from~\eref{deltaL3s} and~\eref{deltaL3} the third-order Noether invariant
\begin{eqnarray*}
\fl I(t,q,\dot q,\ddot q,\dddot q) &= \dot s(t) \frac{\partial L}{\partial\ddot q} - s(t) \frac{\rmd}{\rmd t}\frac{\partial L}{\partial\ddot q}+ s(t)\frac{\partial L}{\partial\dot q} - f(t,q,\dot q)\nonumber\\
&= A(t)\big(\dot s(t)\ddot q -  \ddot s(t)\dot q + \dddot s(t)q - s(t)\dddot q\big)+\dot A(t) \big(\ddot s(t)q - s(t)\ddot q\big)\nonumber\\
&\quad+B(t)\big(s(t)\dot q- \dot s(t) q\big) + \int D(t)s(t)\rmd t.
\end{eqnarray*}

\section{General higher-order linear differential equations}\label{sec:generalisation}

\subsection{Lie symmetry approach}

The reasoning about the Lie linearity symmetry remains evidently valid for an LDE of any order
\begin{equation}
\fl\Delta_n(t,q,q^{(1)},\dots,q^{(n)})=q^{(n)}+a_{n-1}(t)q^{(n-1)}+\dots+a_1(t)q^{(1)}+a_0(t)q+c(t)=0,\label{ODEn}
\end{equation}  
where $q^{(k)}$ designates the $k$-th derivative of $q$. Let $s_1(t),\dots,s_{n-1}(t)$ be independent solutions of the homogeneous equation
\begin{equation}
\fl\Delta_{n\rm h}(t,q,q^{(1)},\dots,q^{(n)})=q^{(n)}+a_{n-1}(t)q^{(n-1)}+\dots+a_1(t)q^{(1)}+a_0(t)q=0.\label{ODEnh}
\end{equation}
Using successively the $n-1$ linearity symmetries, one decreases the order of the LDE~\eref{ODEn} one by one until a first-order LDE. At each step, the intermediate LDE of order $n-k$ is expressed in terms of the derivative of the canonical variable
\begin{equation*}
z_k=\frac{w_k}{\mathscr D_k(t)}
\end{equation*}
where one has introduced the $(k-1)$th-order Wronskian variable
\begin{equation*}
w_k=\left|\begin{array}{ccccc}
s_1(t) & s_2(t) & \dots & s_{k-1}(t) & q \\
s_1^{(1)}(t) & s_2^{(1)}(t) & \dots & s_{k-1}^{(1)}(t) & q^{(1)}  \\
\vdots & \vdots & & \vdots & \vdots \\
s_1^{(k-1)}(t) & s_2^{(k-1)}(t) & \dots & s_{k-1}^{(k-1)}(t) & q^{(k-1)}
\end{array}\right|
\end{equation*}
and where $\mathscr D_k(t)$ is the Wronskian obtained from $w_k$ by substituting $q$ for $s_k(t)$ along the last column. The determinant of a $1\times1$ matrix being conventionally taken as its single coefficient, one has in particular $w_1=q$ and $\mathscr D_1(t)=s_1(t)$. For each integer $k$ comprised between $1$ and $n-1$, the variables $\dot z_k$, $z_{k+1}$ and $w_{k+1}$ are $k$th-order invariants of the $k$-dimensional linearity symmetry group
\begin{equation}
(t,q)\longrightarrow(t,q+\varepsilon_1s_1(t)+\dots+\varepsilon_ks_k(t)).\label{ksymmetry}
\end{equation} 
Differentiating $w_n$ row by row produces the first-order LDE
\begin{equation}
\dot w_n+a_{n-1}(t)w_n+c(t)\mathscr D_{n-1}(t)=0.\label{ODEwn}
\end{equation}
Once integrated, it yields the first integral
\begin{equation*}
I(t,q,q^{(1)},\dots,q^{(n-1)})=w_n\,\rme^{\int a_{n-1}(t)\rmd t}+\int c(t)\mathscr D_{n-1}(t)\rme^{\int a_{n-1}(t)\rmd t}=C.
\end{equation*}
Solving~\eref{ODEwn} for $w_n$ as in paragraph~\ref{preliminary} gives $z_n(t)$. Then, thanks to the formula
\begin{equation}
\dot z_k=\frac{\mathscr D_{k-1}(t)\mathscr D_{k+1}(t)}{\mathscr D_k(t)^2}\,z_{k+1}\label{formula}
\end{equation} 
established in the appendix, $n-1$ quadratures allow to obtain $z_{n-1}(t),\dots, z_1(t)$ and to deduce the general solution $q(t)$. Alternatively, if one knows a last independent solution $s_n(t)$ of~\eref{ODEnh}, one can construct a Cramer system of $n$ first integrals from which one extracts $q(t)$.

\subsection{Noether symmetry approach}

An immediate generalisation of the Lagrangian~\eref{Lagrangian} to the $n$-th order with linear Euler-Lagrange equation 
\begin{equation*}
\mathsf E(L)=\sum_{k=0}^n(-1)^k\frac{\rmd^k}{\rmd t^k}\frac{\partial L}{\partial q^{(k)}}
\end{equation*}
is \textit{a priori} of the form
\begin{equation*}
L=\frac12\sum_{i,j=0}^na_{ij}(t)\,q^{(i)} q^{(j)}+\sum_{i=0}^nb_i(t)q^{(i)}.
\end{equation*}
However, it is a simple task to show, by induction on the non-negative integer $i$, that any term of the form $A(t)q^{(i)}$ can be decomposed as a sum of a term $B(t)q$ and a total derivative. In the same manner, any term of the form $A(t)q^{(i)}q^{(i+j)}$ can be decomposed as a sum of quadratic terms $B_{k}(t)(q^{(k)})^2$ plus a total derivative. All the total derivatives in $L$ can be gauged out and it is sufficient to restrict ourself to a Lagrangian
\begin{equation*}
L=\frac12\sum_{k=0}^n \alpha_k(t)(q^{(k)})^2-\beta(t)q.
\end{equation*}
The corresponding Euler-Lagrange equation of order $2n$ is
\begin{equation*}
\mathsf E(L)=\sum_{k=0}^n(-1)^k\frac{\rmd^k}{\rmd t^k}\Big(\alpha_k(t)q^{(k)}\Big)-\beta(t)=0.
\end{equation*}
Let $s(t)$ be a solution of the homogeneous equation
\begin{equation}
\sum_{k=0}^n(-1)^k\frac{\rmd^k}{\rmd t^k}\Big(\alpha_k(t)q^{(k)}\Big)=\alpha_0(t)q+\sum_{k=1}^n(-1)^k\frac{\rmd^k}{\rmd t^k}\Big(\alpha_k(t)q^{(k)}\Big)=0.\label{equation_s(t)}
\end{equation}
To the first order in $\varepsilon$, it produces the following variation of the Lagrangian:
\begin{equation*}
\delta L=\varepsilon\Bigg[q\,\alpha_0(t)s(t)+
\sum_{k=1}^nq^{(k)}\alpha_k(t)s^{(k)}(t)-\beta(t)s(t)\Bigg].
\end{equation*}
Exploiting relation~\eref{equation_s(t)}, the variation reads
\begin{equation*}
\delta L=\varepsilon\Bigg[-\sum_{k=1}^n(-1)^kq\frac{\rmd^k}{\rmd t^k}\Big(\alpha_k(t)s^{(k)}(t)\Big)+\sum_{k=1}^nq^{(k)}\alpha_k(t)s^{(k)}(t)-\beta(t)s(t)\Bigg].
\end{equation*}
Then, using the identity
\begin{equation}
u(t)v^{(k)}(t)=(-1)^ku^{(k)}(t)v(t)+\frac{\rmd}{\rmd t}\Bigg(\sum_{j=0}^{k-1}(-1)^ju^{(j)}(t)v^{(k-j-1)}(t)\Bigg),\label{derivatives}
\end{equation}
one obtains
\begin{equation}
\delta L=\varepsilon\,\frac{\rmd f}{\rmd t}\label{L1}
\end{equation}
with
\begin{equation*}
f=-\sum_{k=1}^n(-1)^k\sum_{j=0}^{k-1}(-1)^jq^{(j)}\frac{\rmd^{k-j-1}}{\rmd t^{k-j-1}}\Big(\alpha_k(t)s^{(k)}(t)\Big)-\int \beta(t)s(t)\rmd t.
\end{equation*}
Hence, the transformation is definitely a Noether symmetry of $L$. On the other hand, one has
\begin{equation*}
\fl\delta L=\varepsilon\sum_{k=0}^ns^{(k)}(t)\frac{\partial L}{\partial q^{(k)}}=\varepsilon\Bigg\{s(t)\mathsf E(L)
+\frac{\rmd}{\rmd t}\Bigg[\sum_{k=1}^n\sum_{j=0}^{k-1}(-1)^js^{(k-j-1)}(t)\frac{\rmd^j}{\rmd t^j}\Big(\alpha_k(t)q^{(k)}(t)\Big)\Bigg]\Bigg\},
\end{equation*} 
where use has been made of~\eref{derivatives}. This expression together with~\eref{L1} yield the first integral of order $2n-1$
\begin{eqnarray*}
\fl I&=\sum_{k=1}^n\sum_{j=0}^{k-1}(-1)^j\Bigg[s^{(k-j-1)}(t)\frac{\rmd^j}{\rmd t^j}\Big(\alpha_k(t)q^{(k)}(t)\Big)+(-1)^k q^{(j)}\frac{\rmd^{k-j-1}}{\rmd t^{k-j-1}}\Big(\alpha_k(t)s^{(k)}(t)\Big)\Bigg]\\
\fl&\quad+\int\beta(t)s(t)\rmd t.
\end{eqnarray*}

\section{Conclusions}\label{sec:conclusion}

We started this work with an emphasis on the central role played by the linearity symmetry in the treatment of second-order LDEs. In particular, the Lie symmetry allowed the quadrature of the differential equation and the identification of a Wronskian-like Lie invariant. Later the variational approach was pursued by the recognition of the linearity as a Noether symmetry of the Caldirola-Kanai Lagrangian which, in turn, was shown to be a consequence of the Jacobi last multiplier associated with the linearity symmetry. Finally, generalisations of the linearity symmetry to arbitrary orders were investigated. It is hoped that the present work will motivate unified treatments of LDEs based on linearity symmetry principles.

\ack F.~H. acknowledges the support by Con\-se\-lho Na\-cio\-nal de De\-sen\-vol\-vi\-men\-to Cien\-t\'{\i}\-fi\-co e Tec\-no\-l\'o\-gi\-co (CNPq). R.~L. is grateful to the whole Statistical Physics Group and to D.~Malterre for regular stimulating discussions.

\appendix

\section*{Appendix}

In this appendix, we give a proof of formula~\eref{formula}. The variable $z_k$, once expanded in derivatives of $q$, is of the form
\begin{equation*}
z_k=f_{k-1}(t)q^{(k-1)}+f_{k-2}(t)q^{(k-2)}+\dots+f_1(t) q^{(1)}+f_0(t)q.
\end{equation*}
In particular, one has $f_{k-1}(t)=\mathscr D_{k-1}(t)/\mathscr D_k(t)$. Then, the derivative $\dot z_k$ is a certain expression
\begin{equation*}
\dot z_k=\frac{\mathscr D_{k-1}(t)}{\mathscr D_k(t)}\,q^{(k)}+g_{k-1}(t)q^{(k-1)}+\dots+g_1(t)q^{(1)}+g_0(t)q.
\end{equation*}
In the same way, the Wronskian $w_{k+1}$ is of the form
\begin{equation*}
w_{k+1}=\mathscr D_k(t)q^{(k)}+h_{k-1}(t)q^{(k-1)}+\dots+h_1(t)q^{(1)}+h_0(t)q.
\end{equation*}
Consequently, there is a certain relation
\begin{equation*}
\dot z_k-\frac{\mathscr D_{k-1}(t)}{\mathscr D_k(t)^2}\,w_{k+1}=r_{k-1}(t)q^{(k-1)}+\dots+r_1(t)q^{(1)}+r_0(t)q.
\end{equation*}
The left-hand side being an invariant of the symmetry group~\eref{ksymmetry}, the right-hand side is such an invariant as well. That property yields a system of $k$ identities
\begin{equation*}
r_{k-1}(t)s_i^{(k-1)}(t)+\dots+r_1(t)s_i^{(1)}(t)+r_0(t)s_i(t)=0\qquad(i=1,\dots,k)
\end{equation*}
which can be put in the matrix form $S(t)R(t)=0$ with
\begin{equation*}
S(t)=\left(\begin{array}{cccc}
s_1(t)& s_1^{(1)}(t) &\dots & s_1^{(k-1)}(t)\\
s_2(t)& s_2^{(1)}(t) & \dots & s_2^{(k-1)}(t) \\
\vdots & \vdots & & \vdots\\
s_k(t) &  s_k^{(1)}(t)& \dots & s_k^{(k-1)}(t)
\end{array}\right)\quad{\rm and}\quad R(t)=\left(\begin{array}{c}
r_0(t)\\ r_1(t) \\ \vdots \\ r_{k-1}(t)
\end{array}\right).
\end{equation*} 
However, the determinant of $S(t)$, which is equal to $\mathscr D_k(t)$, is nonzero by the assumption on the independence of $s_1(t),\dots,s_k(t)$. Hence, $R(t)$ must vanish identically and one concludes 
\begin{equation*}
\dot z_k=\frac{\mathscr D_{k-1}(t)}{\mathscr D_k(t)^2}\,w_{k+1}=\frac{\mathscr D_{k-1}(t)\mathscr D_{k+1}(t)}{\mathscr D_k(t)^2}\,z_{k+1}\,.
\end{equation*}
\hfill $\Box$

\bibliography{Lienearity_Leone_Haas}

\end{document}